\documentclass[a4paper]{jpconf}
\usepackage{graphicx}
\usepackage{amsmath}	
\usepackage{bm}
\usepackage{color}
\newcommand{\red}[1]{\textcolor{black}{#1}}
\begin{document}
\title{Chiral Symmetry and Electron-Electron Interaction in Many-Body Gap Formation in Graphene}

\author{Y. Hamamoto$^1$, Y. Hatsugai$^{1,2}$, and H. Aoki$^3$}

\address{$^1$ Institute of Physics, University of Tsukuba, Tsukuba 305-8571, Japan}
\address{$^2$ Tsukuba Research Center for Interdisciplinary Materials Science (TIMS), University of Tsukuba, Tsukuba 305-8571, Japan}
\address{$^3$ Department of Physics, University of Tokyo, Hongo, Tokyo 113-0033, Japan}

\ead{hamamoto.yuji.fp@u.tsukuba.ac.jp}

\begin{abstract}
We study a many-body ground state of graphene in perpendicular magnetic fields.  
Chiral symmetry in graphene enables us to determine the many-body ground state, 
which turns out to be a doubly degenerate chiral condensate 
for the half-filled (undoped) case.  
\red{In the ground state a prominent charge accumulation emerges along zigzag edges.}
We also show that gapless excitations are absent despite the presence 
of the robust edge modes,
which is consistent with the Chern number $C=0$.
\end{abstract}
\section{Introduction}
Graphene is a two-dimensional zero-gap semiconductor composed of carbon atoms forming a honeycomb lattice.
Since its discovery~\cite{novoselov04}, there is an upsurge of researches on graphene's intriguing physics, 
which is in may ways intimately linked to a fundamental nature in the lattice structure, i.e., the {\it chiral symmetry}.
For one thing, 
the chiral symmetry guarantees a topological stability of the doubled Dirac cones against finite
perturbations~\cite{hatsugai06}.  We can view this 
as a two-dimensional analogue of the Nielsen-Ninomiya theorem in the four-dimensional lattice gauge theory.
In a magnetic field, the $n=0$ Landau level is unique in terms of the chiral symmetry, where the level remains 
anomalously sharp even in the presence of disorder that 
preserves the chiral symmetry such as ripples~\cite{kawarabayashi09}.

Here we ask a question: How does the chiral symmetry influence {\it many-body physics} of graphene?
There is a prevailing  view that the gap opening in the $n=0$ Landau level 
observed at high magnetic fields~\cite{zhang06}
is due to the electron-electron interaction.
If we take account of the close relation between the chiral symmetry and the $n=0$ Landau level, 
the chiral symmetry may be again playing a crucial role in the many-body 
gap opening.  
In this paper we study the effect of the 
electron-electron interaction in undoped graphene in magnetic fields 
from a viewpoint of the chiral symmetry.

\section{Doubly-degenerate chiral condensate as a many-body ground state}\label{sec:ground}
To model the interacting electrons on a graphene honeycomb lattice,
let us adopt the extended Hubbard model for spinless fermions,
\begin{gather}
 H=H_{\rm kin}+H_{\rm int}
=t\sum_{\langle ij\rangle}(e^{{\rm i}\theta_{ij}}c^\dagger_ic_j+{\rm H.c})+V\sum_{\langle ij\rangle}n_in_j\qquad
(n_i\equiv c^\dagger_ic_i),\label{eq:hamiltonian}
\end{gather}
where electronic spins, assumed to be completely polarized due to the Zeeman effect, are suppressed (hence the on-site interaction is absent).  
Here the screening effect is assumed to be so strong that we can only 
take  the nearest-neighbor repulsion ($V>0$) as 
the electron-electron interaction.  
The kinetic energy part is just the usual hopping, $t$, 
between adjacent sites $\langle ij\rangle$,
and the effect of magnetic field is taken care of by the Pierls phase $\theta_{ij}$.

Since it is difficult to treat the interaction exactly,
we consider a projection onto the $n=0$ Landau level to reduce the Hilbert space,
\red{assuming that the interaction is small compared with the Landau level spacing}.
Note that in the projected subspace one can neglect the kinetic energy.
\red{In the presence of edges, the Landau level projection becomes nontrivial, because quantum Hall edge states appear between Landau levels. We identify the $n=0$ states by checking that the value of charge density away from the edges coincides with that for a system without edges.}

To determine the many-body ground state, 
we can first introduce the chiral operator $\Gamma$ defined as
$\Gamma c_A\Gamma^{-1}=+c_A,\Gamma c_B\Gamma^{-1}=-c_B$ 
with $\Gamma^2=1$,
where $c_{A(B)}$ annihilates an electron at a site in the $A(B)$ sublattice.
$H_{\rm kin}$ has a chiral symmetry, {\it i.e.}, $\{H_{\rm kin},\Gamma\}=0$, 
and, for the one-body, $E=0$ eigenstates $\psi_{E=0}$ one can construct a set of eigenstates of $\Gamma$ as $(1\pm\Gamma)\psi_{E=0}$.  
The even- (odd-)chirality eigenstates are localized on $A(B)$ sublattice.  
If $A$ and $B$ sublattices have the same number of sites,
there are the same number of even- and odd-chirality states.
Thus, in the subspace projected to the $n=0$ Landau level at $E=0$,  
the many-body ground state at half-filling should be a chiral condensate,
which is a state composed totally of $n=0$ states 
polarized to even or odd chirality.
This is because $H_{\rm int}$ should in general have a positive eigenenergy $E_{\rm int}\ge 0$ for a repulsive interaction ($V>0$), 
whereas a chiral condensate exactly has $E_{\rm int}=0$ for 
the nearest-neighbor repulsion.
We have numerically confirmed that this is indeed 
the case, even in the presence of zigzag edges,
through an exact diagonalization of $H_{\rm int}$ for finite systems.  
We stress that the even and odd chiral condensates, 
$|\Psi_{+}\rangle$ and $|\Psi_{-}\rangle$,
constitute a doublet $|\Psi\rangle=(|\Psi_+\rangle,|\Psi_-\rangle)$,
which implies that the ground state is not a simple charge density wave (CDW).
The detail of the doubly degenerate ground state will be discussed elsewhere~\cite{hamamoto}.


\begin{figure}[t]
\begin{minipage}{.27\textwidth}
 \includegraphics[width=\textwidth]{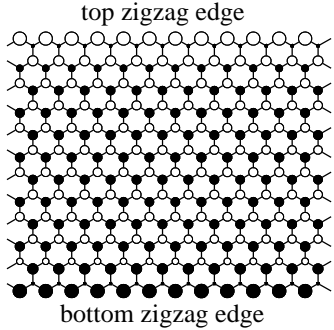} 
\end{minipage}
\hfill
\begin{minipage}{.7\textwidth}
 \caption{\label{fig:zcd}
\red{Local density of states near $E=0$}
in the ground state of a half-filled, 
zigzag-ribbon graphene.  
The system consists of $12\times12$ unit cells,
and the magnetic flux is set to $\phi=1/6$.
$A(B)$ sublattice is represented with solid (open) circles, 
where the size of each circle represents 
a relative value of the
\red{local density of states}.
Contribution from $n<0$ Landau levels is neglected.
A charge accumulation on $A(B)$ sites is seen along the 
top (bottom) zigzag edge, 
while the bulk has a uniform charge distribution over $A$ and $B$.
}
\end{minipage}
\end{figure}

\begin{figure}[t]
\hfill
\begin{minipage}{.45\textwidth}
 \includegraphics[width=\textwidth]{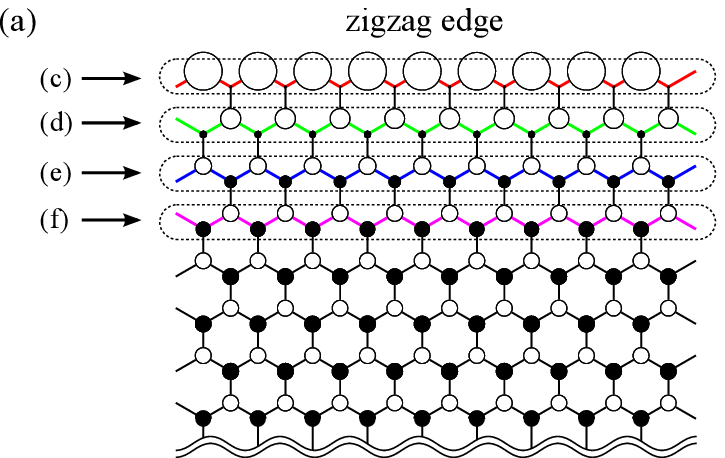}
\end{minipage}
\hfill
\begin{minipage}{.45\textwidth}
 \includegraphics[width=\textwidth]{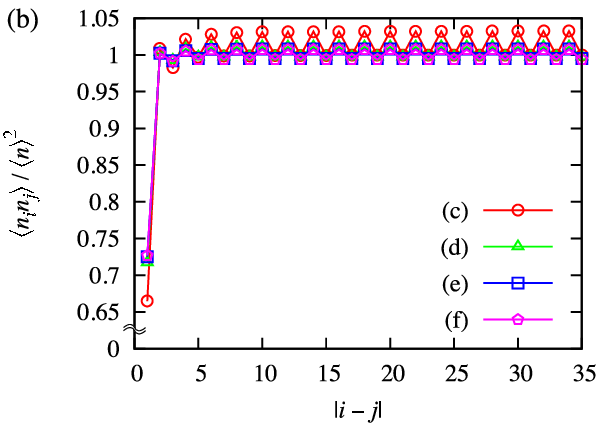}
\end{minipage}
\hfill{}\\
\caption{\label{fig:correlation}
(Color online) (a) For a zigzag ribbon, schematically shown in (a),  
the charge-charge correlation function along the zigzag edge direction 
is displayed \red{in (b)} as a function of distance between two sites 
\red{for several distances from the edge; 0 (c), 1 (d), 2 (e) and 3 (f)}.
The sample has $36\times36$ unit cells with half the 
sample size of 36 (with each unit cell containing two sites),
and the 
magnetic flux is set to $\phi=1/12$.
The correlation function is normalized by square of the bulk charge density
$\langle n\rangle=\red{1/2}$.
}
\end{figure}

\section{\red{Effect of zigzag edges}}
Now that the many-body ground state is determined, 
we can calculate the physical quantities.  
Here we consider a zigzag-edges  ribbon with a 
periodic boundary condition along the ribbon, 
and examine the influence of edges on the ground state.
To see this, we first show the charge distribution in the ground state in a zigzag ribbon%
.
\red{In the ground state with completely filled $n<0$ levels, charge density 
shows no symmetry breaking with a uniform value 1/2.
If we focus on the $n=0$ level, however,
the charge distribution exhibits characteristic patterns
near the edges as shown in Figure \ref{fig:zcd}, which should be observed experimentally as a local density of states (LDOS) around $E=0$ with a scanning tunneling microscope.}
Here the size of a circle denotes the relative value of
\red{LDOS for the $n=0$ level}.
The magnetic flux penetrating through a unit hexagon is set to $\phi=1/6$.
One can see that, near each edge, there is a charge accumulation, 
which is localized on $A$ ($B$) sublattice for the top (bottom) edge.  
In this sense a CDW (or sublattice selected charge accumulation) 
is induced by the edge.  
This behavior reminds us of the presence of one-body edge modes with a flat band near a zigzag edge~\cite{fujita96}.
In the bulk region, on the other hand, 
the CDW decays and the value of LDOS converges to the 
uniform bulk value
.  
It is easy to check that the CDW decays on the magnetic length scale 
$l_B$ away from the edge,
where $l_B\equiv3^{3/4}a/\sqrt{2\pi\phi}\simeq 2a$ 
for the present value of $\phi$ with the lattice spacing $a$.

Next we examine the charge-charge correlation along the edge direction.
In Figure~\ref{fig:correlation}, the two-point correlation function
is plotted as a function of the distance between sites $i$ and $j$.
The correlation function is calculated along zigzag lines enclosed with dashed loops depicted in Fig.2(a), 
and the panels (b) show how the 
charge-charge correlation changes as we go away from the edge.
\red{Sample size dependence is hardly observed between $24\times24$ and $36\times36$ unit cells.}
Directly on the edge~(c), the charge-charge correlation oscillates with a period of two,  indicating realization of the $A$-site selective charging with a long-range order.
The large amplitude of the oscillation reflects sharp accumulation 
of charge right on the edge.
\red{If we go away from the edge into the bulk(d-f),
the oscillation diminishes, but the charge-charge correlation is preserved, 
with a saturation over a distance from the edge as small as 3.
Thus the many-body ground state is shown to exhibit a CDW-like behavior 
over the whole system.
We should stress, however, that the CDW's in the edge and bulk regions are essentially different,
since the former is a consequence of the zigzag edge state localized
on one sublattice, while in the latter the doubly-degenerate ground state still preserves sublattice symmetry.}

This behavior is reminiscent of the $\nu=1/q$ fractional quantum Hall state in two-dimensional electron systems
with an odd $q$.  
In the latter case, the ground state is a quantum liquid with a topological degeneracy $N_D=q$ in the bulk,
whereas the state is locked to an edge state in the edge region.
By contrast, the present ground state is different from the fractional quantum Hall liquid, in the sense that here 
the topological degeneracy $N_D=2$ is a direct consequence of the chiral symmetry, 
and fate of CDW is strongly influenced by the shape of edges.

\section{Chern number}
Finally we examine topological
 properties of the zigzag ribbon in the ground state by calculating the Chern number.
Exact diagonalization of $H_{\rm int}$ gives a finite gap between the ground state
and the first excited states,
so that the Chern number calculated for the ground state is well-defined.
Note that the Chern number should be formulated in 
a matrix (non-Abelian) formalism~\cite{hamamoto},
since the ground state is now doubly degenerate.
Thus we calculate the non-Abelian Chern number~\cite{hatsugai05} as
\begin{gather}
 C=\frac{1}{2\pi i}\int_{\rm BZ} {\rm Tr}[\rmd {\bm A}],\qquad
{\bm A}\equiv\langle\Psi|\rmd\Psi\rangle=\left(
\begin{array}{cc}
 \langle\Psi_+|\rmd\Psi_+\rangle&\langle\Psi_+|\rmd\Psi_-\rangle\\
 \langle\Psi_-|\rmd\Psi_+\rangle&\langle\Psi_-|\rmd\Psi_-\rangle\\
\end{array}\right),
\end{gather}
where ${\bm A}$ is the Berry connection.
The many-body ground state at half-filling turns out to 
have $C=0$,
which indicates that the ground state is a Hall insulator.
While one may have a naive expectation that
the presence of edge states is accompanied by gapless excitations, 
both of the energy spectrum and the Chern number suggest the absence
of gapless (conductive) excitations.  
This may seem contradictory, but is actually 
consistent with an observation that, while a topological state (with 
a nonzero $C$) implies an existence of edge states, 
converse is not necessarily holds.

\section{Summary}
We have studied many-body states in the undoped graphene in a quantum Hall regime, 
and determined the many-body ground state utilizing the chiral symmetry.
The ground state is a doubly degenerate chiral condensate,
where the chiralities of the filled states are fully polarized.
Investigation of charge density and charge-charge correlation has revealed that 
the ground state has charge accumulations with a selected $A$ (or $B$) 
sublattice 
emerging along  zigzag edges.
\red{A CDW-like behavior is observed also in the bulk,
but its origin is essentially different from the edge case.}
Exact diagonalization of $H_{\rm int}$ shows that there is no gapless excitation above the ground state 
despite the presence of the robust edge mode, which 
is consistent with the calculated Chern number of zero.

\ack
The computation in this work has been done using the facilities of the Supercomputer Center,
Institute for Solid State Physics, University of Tokyo. This work was supported in part by
Grants-in-Aid for Scientific Research No. 23340112 and No. 23654128 from the JSPS and No.
22014002, on Priority Areas, from the MEXT.


\section*{References}
\begin{thebibliography}{9}
\bibitem{novoselov04}
Novoselov K S, Geim A K, Morozov S V, Jiang D, Zhang Y, Dubonos S V, Grigorieva I V, Firsov A A
2004 {\it Science} {\bf 306} 666;
Novoselov K S, Geim A K, Morozov S V, Jiang D, Katsnelson M I, Grigorieva I V, Dubonos S V and Firsov A A
2005 {\it Nature} {\bf 438} 197

\bibitem{hatsugai06}
Hatsugai Y, Fukui T and Aoki H 2006 {\it Phys. Rev.} B {\bf 74} 205414

\bibitem{kawarabayashi09}
Kawarabayashi T, Hatsugai Y and Aoki H 2009 {\it Phys. Rev. Lett.} {\bf 103} 156804;
Kawarabayashi T, Morimoto T, Hatsugai Y and Aoki H 2010 {\it Phys. Rev.} B {\bf 82} 195426

\bibitem{zhang06}
Zhang Y, Jiang Z, Small J P. Purewal M S, Tan Y-W, Fazlollahi M, Chudow J D, Jaszczak J A, Stomer H L and Kim P 2006
{\it Phys. Rev. Lett.} {\bf 96} 136806;
Jiang Z, Zhang Y, Stormer H L and Kim P 2007 {\it Phys. Rev. Lett.} {\bf 99} 106802

\bibitem{hamamoto}
Hamamoto Y, Hatsugai Y and Aoki H, unpublished

\bibitem{fujita96}
Fujita M, Wakabayashi K, Nakada K and Kusakabe K 1996 {\it J. Phys. Soc. Jpn.} {\bf 65} 1920

\bibitem{hatsugai05}
Hatsugai Y 2005 {\it J. Phys. Soc. Jpn.} {\bf 74} 1374

\end{thebibliography}
\end{document}